\documentclass[%
 preprint,
 amsmath,amssymb,
prl,
]{revtex4-1}

\usepackage{txfonts}
\usepackage{times}
\usepackage{graphicx}
\usepackage{dcolumn}
\usepackage{bm}

\usepackage{color}
\usepackage[linktocpage,colorlinks=true,linkcolor=blue,citecolor=blue,breaklinks=true]{hyperref}
\usepackage{verbatim}
\usepackage{breakcites}

\usepackage{ulem}

\renewcommand{\vec}[1]{\boldsymbol{#1}}
\newcommand{\tens}[1]{\boldsymbol{#1}}
\newcommand{\bnabla}{\vec{\nabla}}

\begin{document}

\title{Topology and morphology of self-deforming active shells}

\author{Luuk Metselaar}
\affiliation{Rudolf Peierls Centre for Theoretical Physics, Parks Road, Oxford OX1 3PU, United Kingdom}
 
\author{Julia M.\ Yeomans}
 \affiliation{Rudolf Peierls Centre for Theoretical Physics, Parks Road, Oxford OX1 3PU, United Kingdom}
 
\author{Amin Doostmohammadi}
\email{amin.doostmohammadi@physics.ox.ac.uk}
\affiliation{Rudolf Peierls Centre for Theoretical Physics, Parks Road, Oxford OX1 3PU, United Kingdom}


\begin{abstract}
We present a generic framework for modelling three-dimensional deformable shells of active matter that captures the orientational dynamics of the active particles and hydrodynamic interactions on the shell and with the surrounding environment. We find that the cross-talk between the self-induced  flows of active particles and dynamic reshaping of the shell can result in conformations that are tuneable by varying the form and magnitude of active stresses. We further demonstrate and explain how self-induced topological defects in the active layer can direct the morphodynamics of the shell. These findings are relevant to understanding morphological changes during organ development and the design of bio-inspired materials that are capable of self-organisation.
\end{abstract}

\maketitle
\begin{figure*}[t!]
\centering
\includegraphics[width=0.92\linewidth]{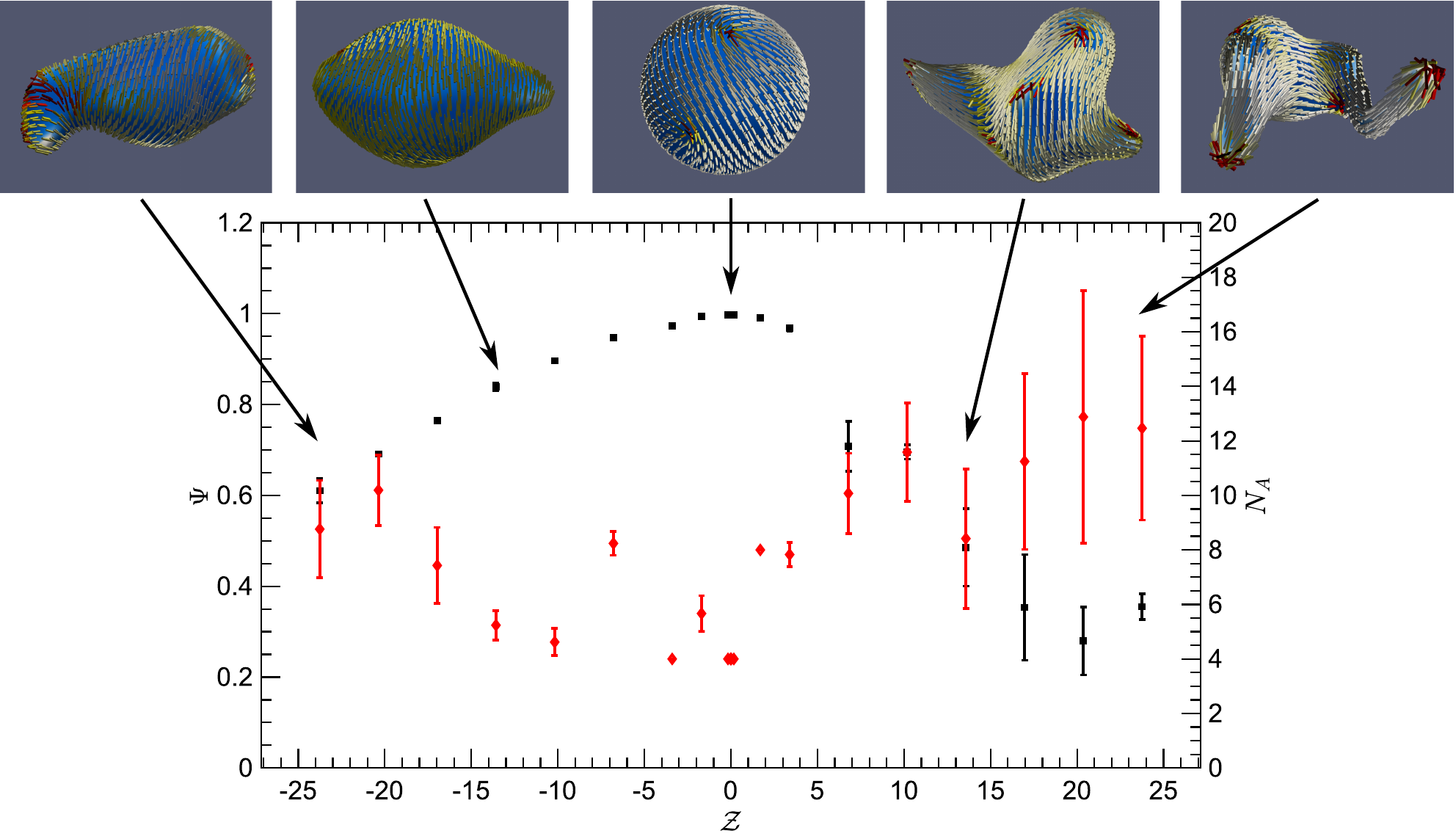}
  \caption{The sphericity $\Psi$ (black squares) and the number of defects $N_A$ (red diamonds) on an active deformable shell as a function of the dimensionless activity $\mathcal{Z}$. For extensile activities ({\it positive} $\mathcal{Z}$) defects are spontaneously created and annihilated at activities where the shell does not yet significantly deform. For contractile activities ({\it negative} $\mathcal{Z}$) the surface becomes less spherical while the defect number remains four. For large extensile and contractile activities protrusions are created. The snapshots indicate representative shapes at different activities. \label{fig:1}
  }
\end{figure*}
The defining feature of living materials such as the cell cytoskeleton, bacterial colonies or cellular tissues is the continuous conversion of chemical energy into mechanical work. This `activity' injects energy at the single particle level by producing active stresses that drive the whole system away from thermodynamic equilibrium~\cite{Ramaswamy2010,Marchetti2013,Cates2015,Doostmohammadi2018}. Importantly, activity is an essential tool for living materials to self-organise and self-assemble into biologically functional systems, and understanding the processes involved may provide biomimetic inspiration for the design of synthetic materials capable of autonomous movement and self-organisation~\cite{Bechinger2016,Needleman2017}.

Dense active materials that produce dipolar flow fields are often well described by continuum, active nematic theories. Instabilities due to the active stresses  destroy any nematic ordering and lead to active turbulence, a state characterised by strong flow vorticity and motile topological defects that are continually created and destroyed. When an active nematic is confined, interplay between the geometrical and topological constraints can lead to a rich dynamical behaviour. For example, an active nematic confined to a spherical shell must, from the Poincar\'e-Hopf theorem, carry a topological charge +2 which can manifest as four +1/2 defects or two +1 defects following intermittent orbits on the surface of the shell. Such defect dynamics has been observed by restricting suspensions of subcellular microtubule filaments driven by kinesin motors to spherical~\cite{Keber2014,Guillamat2018} or toroidal surfaces~\cite{Ellis2018}, and the exotic dynamics has been reproduced in theoretical~\cite{Khoromskaia2017} and particle-based numerical~\cite{Keber2014,Alaimo2017,Henkes2018} studies. Recent extension to ellipsoidal surfaces has investigated the connection between active topological defects and varying surface curvature~\cite{Alaimo2017}. 

While these studies shed light on the motion of active nematics on static shells, in many physiologically relevant conditions  the stresses generated by living systems are capable of actively deforming the surfaces they live on. This is important, in particular, during development, where active stresses not only organise the motion of the cells, but also can tune the morphology of the entire cell assembly to shape it into a particular form ~\cite{Janmey11,Heisenberg13,Guillot13,Lecuit11}. 
Therefore, the dynamics of active self-organisation, eventually, needs to be understood in the context of active self-deformable geometries.

Indeed, a number of recent works have taken first steps in this direction. Encapsulating microtubule/motor protein mixtures within a deformable lipid vesicle, Keber {\it et al.}~\cite{Keber2014} showed that deflating the vesicle can result in tunable dynamic shape changes in the form of ring-shaped, spindle-shaped and anisotropic motile droplets with filipodia-like protrusions. Moreover, Weirich {\it et al.}~\cite{Weirich2019} showed that introducing myosin motors to spindle-shaped droplets of actin filaments results in the formation of contractile stresses at the mid-plane of the droplets that can artificially mimic cell division by splitting the drop into two daughter drops. Miller {\it et al.}~\cite{Miller2018} showed that modeling a deformable shell close to mechanical equilibrium - by separation of chemical and mechanical time-scales - can capture the contraction caused by chemical wave propagation on deformable surfaces and corresponding morphological changes in ascidian and starfish oocytes. Similarly, Mietke {\it et al.}~\cite{Mietke2019} introduced a mechano-chemical coupling to describe active stress organisation and shape changes of axisymmetric surfaces, such as spherical and tubular shells, resulting in shape oscillations and peristaltic motion. 

Notwithstanding these important contributions, modeling active shape-changing surfaces far-from-equilibrium, and beyond axisymmetric shapes, remains challenging. Adding to this complexity, to explain the variety of shape changes observed in recent experiments, accounting for evolution of orientational order, hydrodynamic coupling and the dynamics of topological defects are essential~\cite{Keber2014,Guillamat2018, Ellis2018}. 
Therefore, in this Letter, we present a generic, continuum, three-dimensional framework to study the spatio-temporal dynamics of active self-deforming shells which allows the effects of hydrodynamics and orientational order to be included. We do this by localising a nematic shell at a deformable interface between two (identical) phases of a binary fluid. 

Applying the algorithm to active nematic shells reveals dynamically self-organised morphologies, that can be tuned based on the mechanical properties of the shell, and the magnitude and form (extensile or contractile) of active stress generation. Furthermore, by closely tracking the shape changes and the dynamics of topological defects, we explain the mechanism by which three-dimensional active protrusions are initiated, evolve, and determine the shell morphology.\\

We model a deformable active nematic shell at the interface of two (identical) phases of an isotropic, binary fluid, by adopting the continuum dynamical equations:
\begin{eqnarray}
\bnabla\cdot\bm{u} & = & 0, \label{eq:u0} \\
\rho\left(\partial_t + \bm{u} \cdot \bm{\nabla}\right)\bm{u} & = & -\bnabla p + \bm{\nabla} \cdot (\bm{\Sigma}^\text{passive} + \bm{\Sigma}^\text{active}), \\
\left(\partial_t + \bm{u} \cdot \bm{\nabla}\right)\bm{Q} - \bm{S} & = & \Gamma_Q \bm{H}, \label{eq:Q}\\
\partial_t \phi + \bm{\nabla} \cdot \left(\phi\bm{u}\right) & = & \Gamma_{\phi}  \nabla^2 \mu. \label{eq:phi}
\end{eqnarray}
The fluid flow is given by $\vec{u}$, $\rho$ denotes the fluid density, $p$ the pressure, and $\tens{\Sigma}=\tens{\Sigma}^\text{passive} + \tens{\Sigma}^\text{active}$ is the stress tensor comprising active and passive contributions, as detailed below. 

The phase-field order parameter $\phi$ is used to distinguish the two phases of the binary fluid, and in particular to track the position of the interface between them, which corresponds to the position of the active shell. This approach is similar in spirit to a phase-field formalism that treats coexisting nematic and isotropic fluids which has been used to study active nematic droplets~\cite{Blow14} as model systems for cell motility~\cite{AransonBook,Tjhung17}, or cell division~\cite{Giomi14,Leoni17}, but, as detailed below, we now adapt it to model deformable membranes of an active nematic.

To follow the alignment dynamics of the elongated active particles on the interface, $\tens{Q}$ is defined as the nematic order parameter tensor describing the orientational order of the active particles. In Eq.~(\ref{eq:Q}) the co-rotational term $\bm{S} = \left(\xi \bm{D}+\bm{\Omega}\right)\left(\bm{Q}+\frac{1}{3}\bm{I}\right) + \left(\bm{Q}+\frac{1}{3}\bm{I}\right)\left(\xi \bm{D}-\bm{\Omega}\right)-2\xi\left(\bm{Q}+\frac{1}{3}\bm{I}\right)\text{tr}\left(\bm{Q}\bm{W}\right)$ determines the alignment of the elongated particles in response to gradients in the velocity field that are characterised by rotational $\tens{\Omega}$, extensional $\tens{D}$ and total gradient $\tens{W}$ contributions to the flow, and the flow alignment parameter $\xi$, which is proportional to the aspect ratio of the particles.

 The relaxational dynamics of the nematic tensor $\tens{Q}$ and phase-field order parameter $\phi$ are governed by the molecular field $\bm{H} = -\left(\frac{\delta\mathcal{F}}{\delta\bm{Q}} - \frac{1}{3}\bm{I}\text{tr}\frac{\delta\mathcal{F}}{\delta\bm{Q}}\right)$ and the chemical potential $\mu =  \frac{\delta\mathcal{F}}{\delta\phi}$, respectively. These are determined by minimising a free energy, $\mathcal{F}$. The relaxation strengths are set by the rotational diffusion coefficient $\Gamma_Q$ for the nematic and the mobility $\Gamma_{\phi}$ for the phase field.
\begin{figure*}
\centering
 \includegraphics[width=0.65\linewidth]{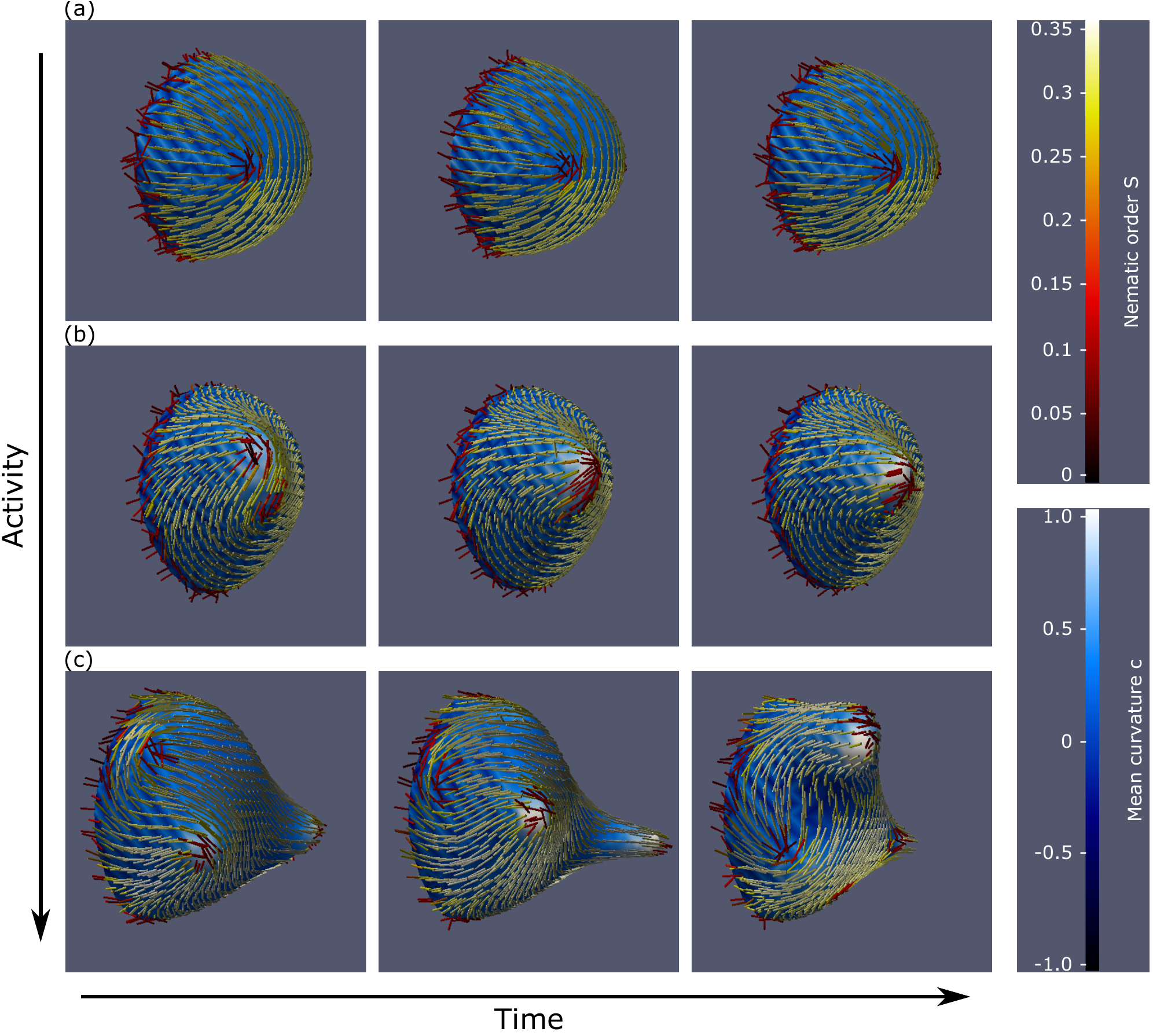}
 \caption{Deformation of an active nematic shell ($R = 12$) attached to a surface, for (a) small activity, $\mathcal{Z} = 0.034$, (b) intermediate activity, $\mathcal{Z} = 8.5$, and (c) large activity, $\mathcal{Z} = 20$. For small activity the two $+1/2$ defects are driven gradually closer together, until the active force is balanced by the elastic force. For intermediate activity the two $+1/2$ defects come into contact and lead to a single protrusion. For large activity motile defects can drive formation of long `tentacles', which will retract again due to surface tension. The directors on the shell are coloured by the magnitude of the order (from red for disordered to yellow to white for fully aligned) and the surface is coloured by the magnitude of the curvature (dark blue for strongly negative to white for strongly positive).
  \label{fig:fig2}}
\end{figure*}

 The first contribution to the free energy is a membrane term, which combines  a mixing free energy corresponding to phase equilibria at $\phi = -1,1$, a bending term  and an interface term:
\begin{equation}
\mathcal{F}_\text{mem} = \frac{\kappa^*}{2}\left(-\phi + \phi^3 - \epsilon^2 \nabla^2\phi\right)^2 + \frac{k_\phi}{2}\left(\bnabla\phi\right)^2.
\label{eq:membrane}
\end{equation} 
$\kappa^*$ is related to the bending rigidity $\kappa$ as $\kappa^* = (4\epsilon^3/3\sqrt{2})\kappa$ and $k_\phi$ is related to the surface tension $\sigma$ by $\sigma \propto \sqrt{k_\phi}$, where
$\epsilon$ characterises the width of the interface.

The orientational order is coupled to the binary order parameter $\phi$ through a bulk free energy
\begin{equation}
\mathcal{F}_\text{b} = A_0\left(\frac{1}{2}\Big(1-\frac{\eta(\phi)}{3}\Big)\text{tr}(\bm{Q}^2) - \frac{\eta(\phi)}{3}\text{tr}(\bm{Q}^3) + \frac{\eta(\phi)}{4}\text{tr}(\bm{Q}^2)^2\right),
\label{eq:fe}
\end{equation}
where $A_0$ is a positive constant. This form of the free energy gives a first order, isotropic-nematic phase transition at $\eta = 2.7$. The key element in writing Eq.~(\ref{eq:fe})
is that, in order to simulate an active nematic at the interface, the expression for $\eta(\phi)$ is chosen as $\eta(\phi) = \eta_0 - \eta_s(\phi-\bar{\phi})^2$. This allows parameters to be chosen such that both free energy minima in $\phi$ correspond to the isotropic phase, but the interface is itself nematic. The bulk free energy is further complemented by the Frank elastic energy $\mathcal{F}_\text{elastic} = \frac{L}{2}\left(\bm{\nabla Q}\right)^2$, penalising orientational deformations, and an interfacial anchoring free energy  $\mathcal{F}_\text{anchoring} =  L_0 \bm{\nabla}\phi \cdot\bm{Q}\cdot \bm{\nabla}\phi$, with $L_0 > 0$ to ensure that the director field lies parallel to the interface.

Using this free energy description, we can write the passive stress tensor in terms of viscous, elastic, and capillary contributions, as in previous work \cite{Metselaar2017}, but now including two additional terms because of the appearance of $\nabla^2 \phi$ in the membrane free energy: $\bnabla \phi \bnabla \frac{\partial \mathcal{F}}{\partial\nabla^2\phi} - \bnabla \bnabla \phi \frac{\partial \mathcal{F}}{\partial\nabla^2\phi}$.
In addition to the passive stresses, the active stress is defined as $\bm{\Sigma}^\text{active} = -\zeta \bm{Q}$ such that gradients in the orientational order $\tens{Q}$ generate active forces that drive active flows. Furthermore, switching the sign of the activity parameter $\zeta$ allows to distinguish extensile $\zeta>0$ from contractile $\zeta<0$ active particles. Equations~(\ref{eq:u0}--\ref{eq:phi})  are solved using the hybrid lattice-Boltzmann method (see S.I. for the choice of the numerical parameters and the corresponding dimensionless variables).

We begin by considering how the morphology of an initially spherical shell of radius $R$ evolves in space and time for different activities $\zeta$. Our main control parameter is the dimensionless number $\mathcal{Z}=\zeta R/\sqrt{\kappa^*k_{\phi}}$, which characterises the ratio of active stresses to the restoring forces due to shell deformation. At small extensile activities the shell remains undeformed, while four $+1/2$ topological defects are present since the total topological charge of the surface must be $+2$. Due to the activity the $+1/2$ defects orbit the surface, reproducing the experimental observations of Keber {\it et al.}~\cite{Keber2014}.  Increasing activity, however, results in strong enough active stresses to deform the shell, creating an autonomously shape-changing material. To quantify the deviation of the shell morphology from spherical we calculate the sphericity $\Psi = 36\pi {V_s}^2/{A_s}^3$,  where $V_s$ is the shell volume and $A_s$ is its surface area. 

As the extensile activity is increased beyond a certain threshold, the sphericity $\Psi$ drops below one, indicating that the initial spherical shell is self-developing into a more anisotropic morphology (see Fig.~\ref{fig:1}; {\it black squares}). At the same time, monitoring the average number of $+1/2$ topological defects on the surface (see Fig.~\ref{fig:1}; {\it red diamonds}) shows that this increases from four.
Since the total topological charge on the surface of the shell has to remain $+2$ this indicates that pairs of $\pm 1/2$ topological defects are nucleated and the periodic patterns of the defect motion have now transitioned into active turbulence on the shell surface. 

Although for extensile activities the shell deformations predominantly occur for active stresses that are strong enough to prompt defect pair nucleation and establish active turbulence, shape changes in contractile systems are possible even without the nucleation of defect pairs. At small contractile activities (Fig.~\ref{fig:1}; {\it negative $\mathcal{Z}$}-values) two $+1/2$ defects localise at each pole and, unlike in extensile systems, self-propel towards their comet-like tails, stretching the initially spherical shell into a spindle shape. The spindle configuration of the shells  resembles the tactoids formed in lyotropic liquid crystals~\cite{Kim13,Metselaar2017,Genkin18} and in recently reported droplets of actin filaments~\cite{Weirich2019}. 
Increasing the contractile activity first further elongates the spindles. Then, as it is increased still further, pairs of $\pm 1/2$ topological defects start to nucleate on the surface, which leads to the formation of protrusions and troughs on the shell and results in the emergence of more complex morphologies.

A common feature observed in both extensile and contractile systems  at high activities is the emergence of protrusions and troughs on the shell, which appear to be closely connected to the dynamics of topological defects. To test this interconnection, we next simulate a deformable shell in a more constrained setup: a hemisphere fixed on a substrate. We initialise a hemispherical, extensile active nematic shell in an isotropic fluid background. The active nematic particles are homeotropically anchored to the underlying substrate, such that the total topological charge of the nematic on the half-sphere is  $+1$. The membrane is then allowed to deform continuously, but the interaction with the substrate constrains the deformation. 

\begin{figure}
\centering
 \includegraphics[width=0.95\linewidth]{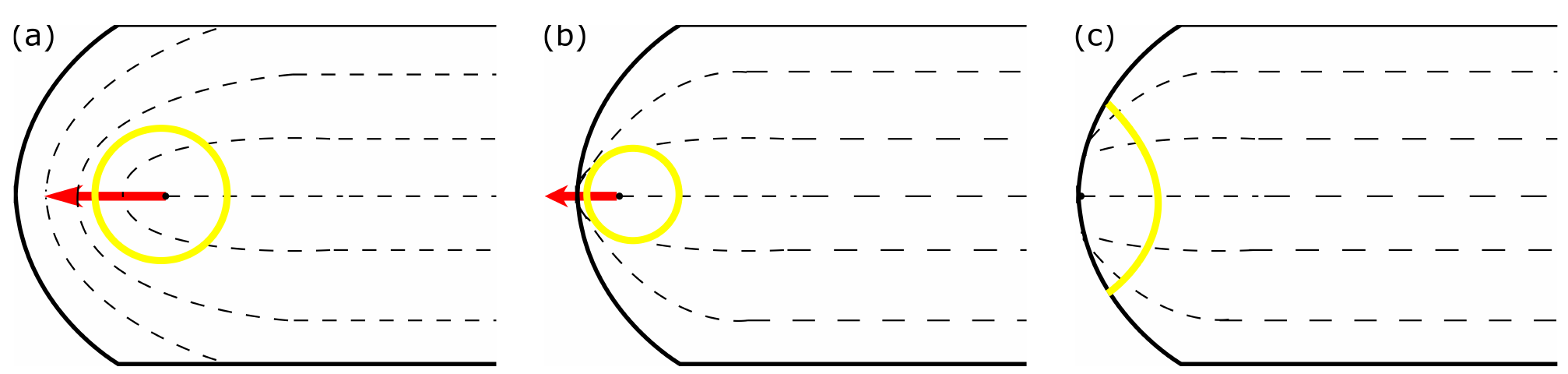}
  \caption{Schematic of a $+1/2$ topological defect approaching the tip of a protrusion (projection onto the plane). (a) The motile defect is moving over the surface of the protrusion, causing it to grow. (b) Closer to the tip the gradients will become smaller and the active stress will decrease. (c) When the defect has reached the tip of the protrusion, calculating the winding number along the yellow line gives a topological charge of $+1$. The gradients in the director field are small, and the active flow will therefore be small as well.\label{fig:schematic}}
\end{figure}

Fig.~\ref{fig:fig2}(a) shows the deformation of the nematic hemisphere at small activity. The two $+1/2$ defects are driven towards each other, until the elastic repulsion keeps them at a fixed distance. The activity is not large enough to create additional defect pairs, so the configuration is in an unstable steady state. For intermediate activity the motile defects are driven together and merge into a single $+1$ defect (Fig.~\ref{fig:fig2}(b)). The stresses at the  $+1$ defect create a single protrusion, indicated by the region of large curvature in Fig.~\ref{fig:fig2}(b). Indeed when the curvature is sufficiently large, the geometry itself will aid the stability of a $+1$ defect (see Fig.~\ref{fig:schematic} for a schematic drawing).  

\begin{figure}
\centering
 \includegraphics[width=0.9\linewidth]{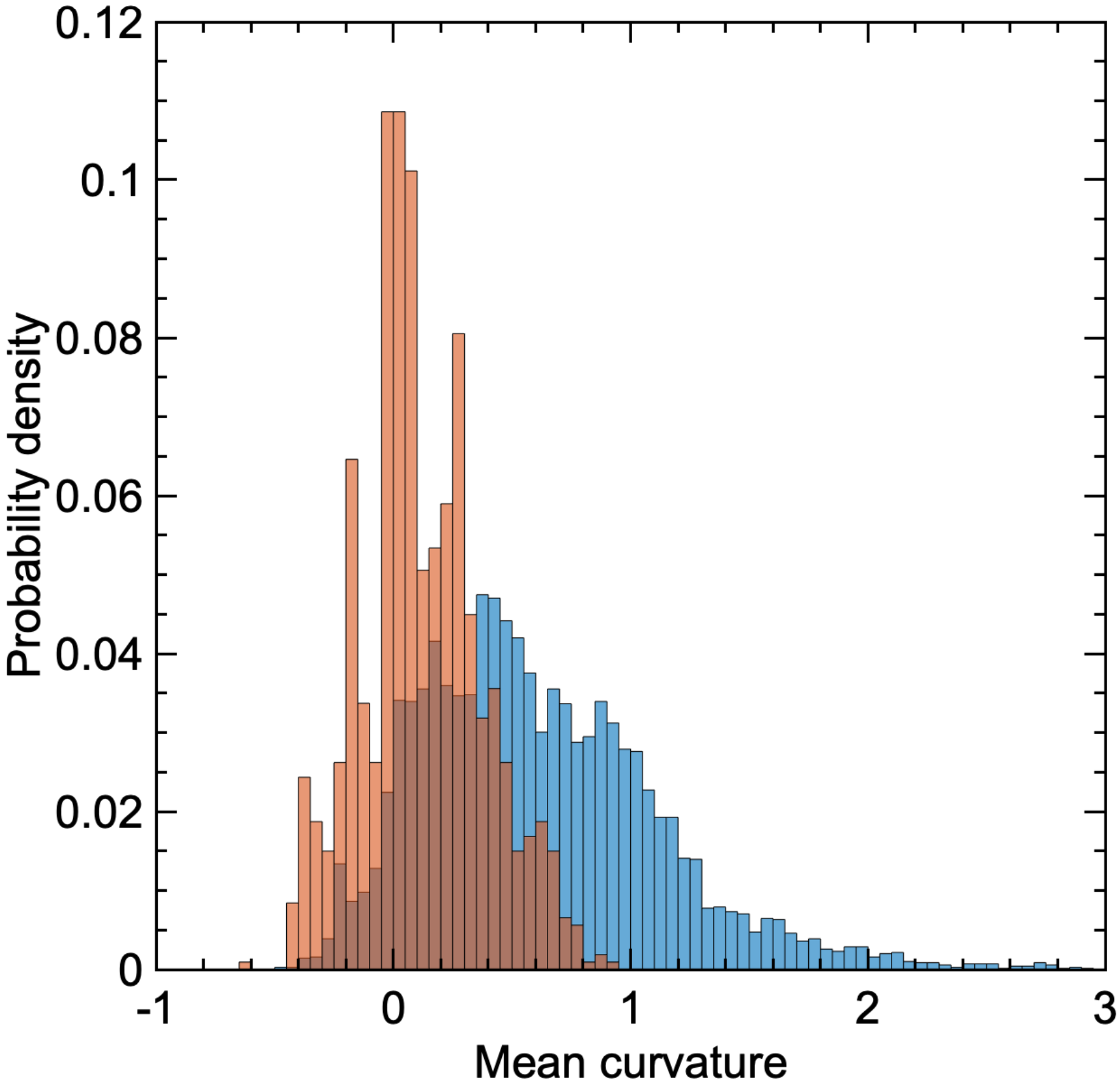}
  \caption{Normalised histogram of the mean curvature for negative (red) and positive (blue) topological charges on a half-sphere at $\mathcal{Z} = 20$. Negative topological charges are more likely found in regions of small mean curvature, whereas positive topological charges move towards and generate their own strongly curved surfaces. \label{fig:histogram}}
\end{figure}

For large activity the dynamics on the surface of the shell is much more chaotic, and additional defect pairs form and annihilate. It is energetically favourable for +1/2 (-1/2) topological defects to lie in regions of larger (smaller) mean curvature.
Because of their motility, topological defects in active systems can move to such favourable locations. Here, because the surface is deformable they can also dynamically drive variations in the curvature. To quantify the interconnection between the topological defects and the surface curvature, we measure the histogram of the mean curvature for positive and negative topological charges (Fig.~\ref{fig:histogram}).
Negative topological charges are clearly more likely to be found in regions of small mean curvature, whereas positive topological charges move towards regions with large mean curvature and generate their own strongly curved surfaces.

Remarkably, the flows created by motile $+1/2$ defects can drive the formation of long ``tentacles'' (Fig.~\ref{fig:fig2}(c)). To conserve charge, when a $+1/2$ defect reaches the tip of a protrusion and becomes a $+1$ topological defect, it has to leave behind a $-1/2$ topological defect, which generates a trough on the surface of the shell (centre of Fig.~\ref{fig:fig2}(c)). With the $+1$ defect on the tip of the protrusion, the active flow becomes negligible in the tentacle, since the gradient of $\tens{Q}$ and, consequently, the active force $f_{\text{active}} = -\zeta\vec{\nabla}\cdot\tens{Q}$ vanish at the tip (Fig.~\ref{fig:schematic}). Surface tension will then lead to retraction of the protrusion.\\


To summarise, we have introduced a way to simulate deformable, active nematic shells. Our results demonstrate that the active flows associated with gradients of the nematic tensor play a key role in dictating the shell dynamics and morphology. In particular, self-motile +1/2 topological defects can drive the formation of long, tentacle-like protrusions analogous to those observed experimentally in flexible vesicles coated with suspensions of microtubules and motor proteins~\cite{Keber2014} and reminiscent of the tentacles of the multicellular polyp Hydra~\cite{Livshits17,Braun18}.

\section{Acknowledgement}
We thank Kinneret Keren, Rian Hughes, and Kristian Thijssen for helpful discussions. This project has received funding from the European Union's Horizon 2020 research and innovation programme under the DiStruc Marie Sklodowska-Curie Grant Agreement No. 641839. A.D. was supported by a Royal Commission for the Exhibition of 1851 Research Fellowship.

\bibliography{refs.bib}
\end{document}